\documentclass[conference, a4paper]{IEEEtran}

\usepackage[turkish]{babel}
\usepackage[utf8]{inputenc}
\usepackage[T1]{fontenc}

\usepackage[cmex10]{amsmath}
\usepackage{amssymb}

\usepackage{graphicx}

\usepackage{multirow}
\usepackage{array}
\usepackage{pifont}
\usepackage{cite}
\usepackage[caption=false,lofdepth,lotdepth]{subfig}
\usepackage{setspace}

\usepackage[hidelinks]{hyperref}

\AtBeginDocument{\renewcommand\tablename{Table}}
\AtBeginDocument{\renewcommand\figurename{Figure}}
\AtBeginDocument{\renewcommand\abstractname{Abstract}}
\AtBeginDocument{\renewcommand\refname{References}}

\begin{document}

\title{Investigating the Potential of Multi-Stage Score Fusion in Spoofing-Aware Speaker Verification}

% author names and affiliations
% use a multiple column layout for up to three different
% affiliations
\author{%
\IEEEauthorblockN{Oğuzhan KURNAZ}
\IEEEauthorblockA{Mechatronics Engineering\\
Bursa Technical University\\
Bursa, Türkiye\\
oguzhan.kurnaz@btu.edu.tr}
\and
\IEEEauthorblockN{Tomi H. KINNUNEN}
\IEEEauthorblockA{School of Computing\\
University of Eastern Finland\\
Joensuu, Finland\\
tomi.kinnunen@uef.fi}
\and
\IEEEauthorblockN{Cemal HANİLÇİ}
\IEEEauthorblockA{Electric and Electronics Engineering\\
Bursa Technical University\\
Bursa, Türkiye\\
cemal.hanilci@btu.edu.tr}
}

\maketitle

% % make the title area
% \IEEEoverridecommandlockouts
% \maketitle
% \IEEEpubid{%
% 	\makebox[\columnwidth]{\textbf{979-8-3315-6655-5/25/\$31.00~\copyright~2025 IEEE}\hfill}%
% 	\hspace{\columnsep}%
% 	\makebox[\columnwidth]{}%
% }

% % -- Hemen \maketitle'dan sonra:
% \hfill{%
% \hspace{\columnsep}%
% \makebox[\columnwidth]{}%
% }

\begin{abstract}
Despite improvements in automatic speaker verification (ASV), vulnerability against spoofing attacks remains a major concern. In this study, we investigate the integration of ASV and countermeasure (CM) subsystems into a modular spoof-aware speaker verification (SASV) framework. Unlike conventional single-stage score-level fusion methods, we explore the potential of a multi-stage approach that utilizes the ASV and CM systems in multiple stages. By leveraging ECAPA-TDNN (ASV) and AASIST (CM) subsystems, we consider support vector machine and logistic regression classifiers to achieve SASV. In the second stage, we integrate their outputs with the original score to revise fusion back-end classifiers. Additionally, we incorporate another auxiliary score from RawGAT (CM) to further enhance our SASV framework. Our approach yields an equal error rate (EER) of \emph{$1.30\%$} on the evaluation dataset of the SASV2022 challenge, representing a $24\%$ relative improvement over the baseline system.
%\boldmath
\end{abstract}

% \IEEEpubidadjcol 

\section{Introduction}

ASV systems compare enrollment and test utterances to verify speaker identity, and recent advances in deep speaker embeddings have significantly improved performance \cite{bai2021speaker, Desplanques_2020, 8461375}. Despite this, ASV systems remain vulnerable to spoofing attacks \cite{Shim2022}, prompting the development of various countermeasures (CMs). While ASV and CM were traditionally studied separately, recent studies integrate them within spoof-aware speaker verification (SASV) systems \cite{Shim2022} to improve robustness.

To address spoofing, two main SASV design paradigms have emerged: \emph{(i)} modular systems with separately developed ASV and CM components \cite{Zhang2022}, and \emph{(ii)} monolithic end-to-end architectures \cite{Shim2022, kang22_interspeech}. This work focuses on the former due to its transparency and flexibility for subsystem updates.

One key design question is how to fuse ASV and CM systems. The literature identifies three strategies: \emph{decision-level} fusion via binary logic \cite{Wu2022e}, \emph{embedding-level} fusion of feature vectors \cite{Ge2022}, and \emph{score-level} fusion using soft outputs \cite{Shim2022}. Score fusion is especially attractive for its simplicity, and recent studies \cite{Wu2022e, Wang2022b} have shown its effectiveness in combining ASV and CM scores.

While conventional score-level fusion is widely used in SASV, it has notable limitations. It combines ASV and CM scores without fine-tuning, leading to potential information loss and limited adaptability across datasets due to differing data characteristics \cite{Zhang2022}. In contrast, multi-stage fusion merges scores across successive stages, better modeling subsystem interactions and improving robustness \cite{DBLP:journals/corr/abs-2106-15793}. This approach has shown success in other domains like emotion recognition \cite{Atmaja_2020} and face matching \cite{9522754}, though its potential in SASV remains unexplored. To fill this gap, we propose a modular multi-stage score-level fusion method, enabling refined integration of ASV and CM outputs via classifiers such as SVM or logistic regression.
\begin{figure*}[h]
    \centering
    \begin{tabular}{cc}
        \shorthandoff{=}
        \includegraphics[scale=0.5]{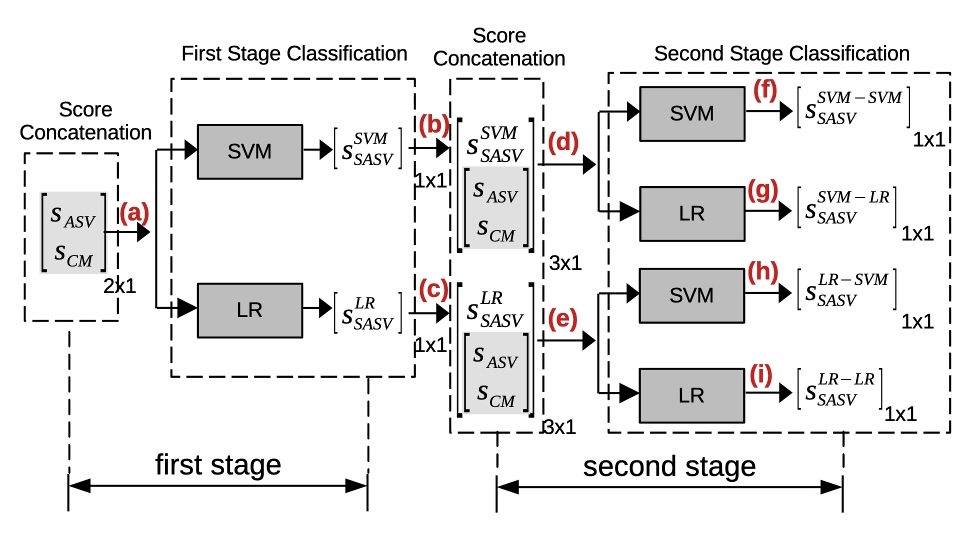} & 
        \shorthandon{=}
        \shorthandoff{=}
        \includegraphics[scale=0.5]{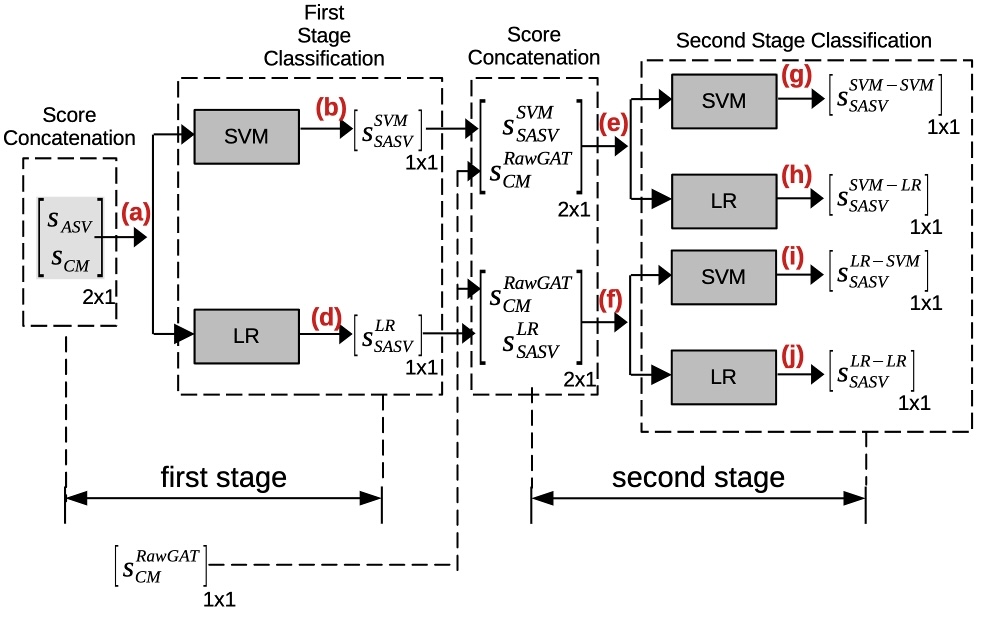} \\
        \shorthandon{=}
        (a) Self-Augmented Multi-Stage Fusion &
        (b) Externally-Augmented Multi-Stage Fusion \\
    \end{tabular}
    \caption{The proposed multi-stage fusion methods: (a) Self-Augmented and (b) Externally-Augmented. 
    In (a), the scores from ECAPA-TDNN and AASIST are combined and sent to the initial classification block. 
    SVM and LR classifiers produce two outputs, which are then merged with the initial scores. 
    Subsequently, the SVM and LR classifiers generate four scores using two different three-dimensional scores. 
    In (b), the second score concatenation block combines the classifier scores from the previous block 
    with new scores extracted from the RawGAT model. The rest of the process is similar to the self-augmented version. 
    Red letters indicate the pathway for score generation in both figures.}
    \label{fig:fig9}
    \vspace{-0.2cm}
\end{figure*}

\section{Related Work}

In the SASV2022 challenge baselines, both score- and embedding-level fusion techniques were explored \cite{Shim2022}. For score fusion, ASV and CM scores from ECAPA-TDNN and AASIST were aggregated. For embedding fusion, enrollment and test embeddings from the ASV system and test embeddings from the CM system were concatenated and passed through a deep neural network to compute SASV scores. In \cite{Zhang2022}, a probabilistic fusion framework was introduced to integrate ASV and CM scores, supporting both direct inference and fine-tuning. Various score-level fusion and cascade systems were proposed in \cite{Wang2022b}, while \cite{Wu2022e} introduced a multi-level fusion system combining ASV and CM scores in a prediction layer. Most of these studies focus on score-level fusion using summation, multiplication, or single-stage methods. However, multi-stage score fusion remains largely unexplored in SASV. Motivated by this gap, we propose a novel multi-stage score-level fusion approach for SASV.

\vspace{-0.25cm}
\section{Methods}
\subsection{Traditional Single-Stage Fusion}
\vspace{-0.15cm}
Traditional score fusion methods are widely used to combine outputs from multiple subsystems, typically through simple operations like score summation, multiplication, or concatenation. These fused scores are often processed in a two-stage pipeline, where a back-end classifier is used for final decision-making. In score summation, subsystem scores are directly added to produce a merged score, while score multiplication combines them by multiplying the individual scores.

\begin{figure*}[h]
    \centering
    \begin{tabular}{@{\hspace{-0.7cm}}ccc@{}}
        \shorthandoff{=}
        \includegraphics[width=0.32\textwidth]{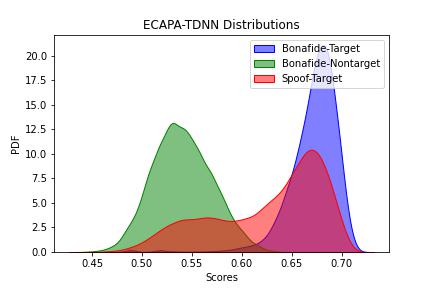} &
        \shorthandon{=}
        \shorthandoff{=}
        \includegraphics[width=0.32\textwidth]{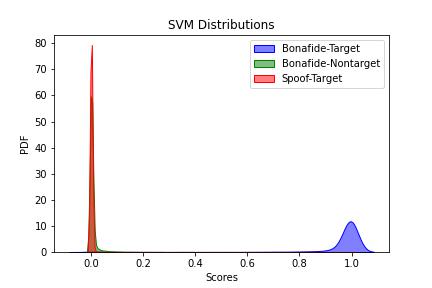} &
        \shorthandon{=}
        \shorthandoff{=}
        \includegraphics[width=0.32\textwidth]{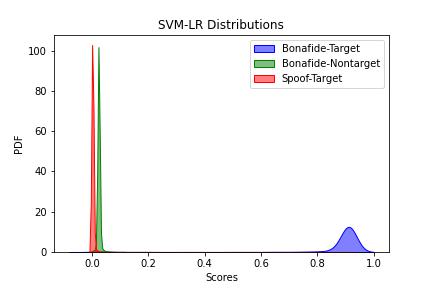} \\
        \shorthandon{=}
        (a) ECAPA-TDNN & (b) Single-stage SVM & (c) Multi-stage SVM-LR \\ 
    \end{tabular}
    \caption{From left to right, the score distributions of (a) ECAPA-TDNN, (b) single-stage SVM classifier with three scores, and (c) multi-stage LR-SVM ( (a)-(b)-(d)-(g) ) with three scores.}
    
    \label{fig:fig10}
    \vspace{-0.5cm}
\end{figure*}

\subsection{Self-Augmented Fusion Method}
\vspace{-0.1cm}

The \emph{self-augmented} multi-stage fusion approach, illustrated in Fig. \ref{fig:fig9} (a), integrates ASV and CM scores from ECAPA-TDNN and AASIST, respectively. A classifier is first trained on these scores to generate initial SASV scores. These are then concatenated with the original ASV and CM scores and passed to a second classifier. This framework supports two variants: \emph{multi-stage fusion with two scores} and \emph{multi-stage fusion with three scores}. In the two-scores method, ASV (from ECAPA-TDNN) and CM (from AASIST) scores are fused with the SASV scores. The three-scores method follows the same structure but incorporates an additional CM score from the RawGAT model, resulting in three initial inputs. The formulation for the two-scores method is given as follows:

\vspace{-0.3cm}
\begin{equation}
\footnotesize
\begin{split}
    s = &\mathrm{C_{2}^{j}} \Bigl( \bigr[ \mathrm{C_{1}^{i}} \Bigl( \bigr[ s_{\texttt{ASV}}^\texttt{{ECAPA}},s_{\texttt{CM}}^\texttt{{AASIST}} \bigr] \Bigl), s_{\texttt{ASV}}^\texttt{{ECAPA}},s_{\texttt{CM}}^\texttt{{AASIST}} \bigr] \Bigl) \\
    &i, j \in \bigl\{ SVM, LR \bigl\}
\end{split}
\label{eq:mssf}
\end{equation}

Similarly, the self-augmented multi-stage fusion with three scores is expressed as follows:

\vspace{-0.3cm}
\begin{equation}
\footnotesize
\begin{split}
     s = &\mathrm{C_{2}^{j}} \Bigl( \bigr[ \mathrm{C_{1}^{i}} \Bigl( \bigr[ s_{\texttt{ASV}}^\texttt{{ECAPA}},s_{\texttt{CM}}^\texttt{{AASIST}},s_{\texttt{CM}}^\texttt{{RawGAT}} \bigr] \Bigl), s_{\texttt{ASV}}^\texttt{{ECAPA}},s_{\texttt{CM}}^\texttt{{AASIST}}, s_{\texttt{CM}}^\texttt{{RawGAT}} \bigr] \Bigl) \\
    &i, j \in \bigl\{ SVM, LR \bigl\}
\end{split}
\label{eq:mssf}
\end{equation}

\noindent Here, $s_{\texttt{ASV}}^\texttt{{ECAPA}}$ and $s_{\texttt{CM}}^\texttt{{AASIST}}$ represent the initial ASV and CM scores, respectively. $\mathrm{C_{1}^{i}}$ and $\mathrm{C_{2}^{j}}$ denote the first and second stage classifiers, with $i$ and $j$ being SVM or LR classifiers.
\vspace{0.15cm}

According to choice of the fusion models (SVM or LR) in the first and second stages of the self-augmented method, four different two-stage fusion strategies are considered, as illustrated in Fig. \ref{fig:fig9} (a).

\begin{table*}[h!]
\centering
\caption{\label{table2}{\it Results of multi-stage score-level fusion. 
The “Stage” column indicates the source stage of each score, while 
“Fusion Type” specifies the applied fusion strategy. The “System” column 
refers to the system used. In the multi-stage section, four-letter 
combinations correspond to score generation paths illustrated in 
Figs.~\ref{fig:fig9}(a) and \ref{fig:fig9}(b). The “Classifiers” column 
identifies the classifiers employed in those figures. Finally, the “Features” 
column lists the features used, where E, A, and R represent scores from 
the ECAPA-TDNN, AASIST, and RawGAT models, respectively.}}
\begin{spacing}{1.2}

\scriptsize
\begin{tabular}{|c|c|c|c|c|c|c|c|c|c|c|c|c|}
\hline
Stage & Fusion Type & System & Classifiers & Scores & \multicolumn{2}{c|}{SASV-EER (\%)} & \multicolumn{2}{c|}{SV-EER (\%)} & \multicolumn{2}{c|}{SPF-EER (\%)} & \multicolumn{2}{c|}{a-DCF} \\
\cline{6-13}
& & & & & Dev & Eval & Dev & Eval & Dev & Eval & Dev & Eval \\
\hline \hline
- & Score sum & Baseline B1\cite{Shim2022} & 
- & E, A & 1.01 & 1.71 & 1.99 & \textbf{1.66} & 0.23 & 1.76 & N/A & N/A \\
First & Single-stage & (a)-LR & $C_{1}^{LR}$ & E, A & 1.15 & 1.61 & 2.02 & 1.86 & 0.25 & 1.26 & 0.030 & 0.036 \\
First & Single-stage & (a)-SVM & $C_{1}^{SVM}$ & E, A & \textbf{0.94} & 1.60 & \textbf{1.82} & 1.69 & 0.25 & 1.47 & \textbf{0.027} & 0.038 \\
First & Single-stage & Gaussian back-end & $C_{1}^{Gaus}$ & E, A & 1.08 & 1.60 & 2.01 & 1.99 & 0.07 & 0.81 & 0.33 & 0.032 \\
\hline \hline
First & Single-stage & (a)-LR & $C_{1}^{LR}$ & E, A, R & 1.08 & 1.55 & 1.89 & 1.75 & 0.20 & 1.30 & 0.029 & 0.033 \\
First & Single-stage & (a)-SVM & $C_{1}^{SVM}$ & E, A, R & \textbf{0.94} & 1.38 & 1.75 & 1.68 & 0.13 & 0.93 & \textbf{0.027} & 0.029 \\
\hline \hline
Second & Self-Aug. Multi-stage & (a)-(c)-(e)-(i) & $C_{2}^{LR}(C_{1}^{LR})$ & E, A & 1.17 & 1.58 & 2.02 & 1.94 & \textbf{0.07} & 1.02 & 0.031 & 0.035 \\
Second & Self-Aug. Multi-stage & (a)-(c)-(e)-(h) & $C_{2}^{SVM}(C_{1}^{LR})$ & E, A & 1.82 & 2.36 & 2.22 & 2.40 & 1.82 & 2.36 & 0.034 & 0.034 \\
Second & Self-Aug. Multi-stage & (a)-(b)-(d)-(g) & $C_{2}^{LR}(C_{1}^{SVM})$ & E, A & 1.00 & 1.60 & 1.95 & 1.73 & \textbf{0.07} & 1.23 & 0.028 & 0.037 \\
Second & Self-Aug. Multi-stage & (a)-(b)-(d)-(f) & $C_{2}^{SVM}(C_{1}^{SVM})$ & E, A & 2.49 & 2.53 & 2.49 & 2.53 & 2.29 & 2.36 & 0.032 & 0.032 \\
\hline \hline
Second & Self-Aug. Multi-stage & (a)-(c)-(e)-(i) & $C_{2}^{LR}(C_{1}^{LR})$ & E, A, R & 1.08 & 1.43 & 1.95 & 1.76 & 0.22 & 0.69 & 0.029 & 0.031 \\
Second & Self-Aug. Multi-stage & (a)-(c)-(e)-(h) & $C_{2}^{SVM}(C_{1}^{LR})$ & E, A, R & 2.09 & 2.09 & 2.16 & 2.14 & 1.75 & 1.69 & 0.030 & 0.030 \\
Second & Self-Aug. Multi-stage & (a)-(b)-(d)-(g) & $C_{2}^{LR}(C_{1}^{SVM})$ & E, A, R & 0.95 & \textbf{1.30} & \textbf{1.82} & 1.68 & 0.20 & \textbf{0.59} & \textbf{0.027} & \textbf{0.028} \\
Second & Self-Aug. Multi-stage & (a)-(b)-(d)-(f) & $C_{2}^{SVM}(C_{1}^{SVM})$ & E, A, R & 2.16 & 2.03 & 2.29 & 2.09 & 1.68 & 1.58 & 0.029 & \textbf{0.028} \\
\hline \hline
Second & Ext-Aug. Multi-stage & (a)-(d)-(f)-(j) & $C_{2}^{LR}(C_{1}^{LR})$ & E, A, R & 1.15 & 1.53 & 2.02 & 1.88 & 0.27 & 0.76 & 0.030 & 0.031 \\
Second & Ext-Aug. Multi-stage & (a)-(d)-(f)-(i) & $C_{2}^{SVM}(C_{1}^{LR})$ & E, A, R & 1.82 & 1.96 & 2.02 & 1.84 & 1.82 & 2.01 & 0.031 & 0.029 \\
Second & Ext-Aug. Multi-stage & (a)-(b)-(e)-(h) & $C_{2}^{LR}(C_{1}^{SVM})$ & E, A, R & 0.94 & 1.53 & 1.82 & 1.70 & 0.27 & 1.07 & \textbf{0.027} & 0.035 \\
Second & Ext-Aug. Multi-stage & (a)-(b)-(e)-(g) & $C_{2}^{SVM}(C_{1}^{SVM})$ & E, A, R & 2.36 & 2.33 & 2.36 & 2.36 & 2.36 & 2.29 & 0.030 & 0.029 \\
\hline
\end{tabular}
\end{spacing}
\end{table*}

\begin{table*}[h!]
\caption{\label{table3} {\it Per-attacks scores. In this table, the "System" column indicates the system utilized. The columns between A07-A19 show the attack types and the pooled column shows that all attack types are pooled. The best results of every method in Table \ref{table2} are shown.}}
\scriptsize
\centerline{
\begin{tabular}{|c|c|c|c|c|c|c|c|c|c|c|c|c|c|c|}
\hline
System  &   A07 &   A08 &   A09 &   A10 &   A11 &   A12 &   A13 &   A14 &   A15 &   A16 &   A17 &   A18 &   A19 &   Pooled \\
\hline  \hline
ECAPA-TDNN  & 32.66 & 18.80  &  2.20  & 50.61 & 47.08 & 39.56 & 11.62 & 35.39 & 36.54 & 60.71 &  1.85 &  2.38 &  4.77 & 30.75 \\
AASIST &  0.80  &  0.44 &  \textbf{0.00}    &  \textbf{1.06} &  0.31 &  0.91 &  0.10  &  0.14 &  0.65 &  \textbf{0.72} &  1.52 &  3.40  &  0.62 & 1.13 \\
Baseline-B1 &  2.05 &  0.69 &  0.07 &  7.19 &  0.32 &  4.42 &  \textbf{0.07} &  0.08 &  1.75 &  1.18 &  0.73 &  1.18 &  0.63 & 0.78 \\ 
(a)-SVM (with 3 scores) &  0.78 &  0.65 &  0.04 &  1.62 &  0.19 &  1.12 & 0.10 &  0.08 &  0.80 &  1.20 &  0.88 &  1.49 &  0.73 & 0.93 \\
(a)-(c)-(e)-(i) (Self-Aug. with 2 scores) &  0.95 &  0.35 &  \textbf{0.00}  &  2.95 &  0.11 &  1.73 &  \textbf{0.07} &  0.08 &  0.92 &  0.96 &  0.58 &  0.99 &  0.41 &  1.02 \\
(a)-(b)-(d)-(g)  (Self-Aug. with 3 scores) &  \textbf{0.26} &  \textbf{0.34} &  \textbf{0.00}    &  1.14 &  \textbf{0.07} &  \textbf{0.45} &  \textbf{0.07} &  \textbf{0.06} &  \textbf{0.31} &  1.00    &  \textbf{0.51} &  1.21 &  0.39 & \textbf{0.59}  \\
(a)-(d)-(f)-(j) (Ext-Aug. with 3 scores)  &  0.33 &  0.37 &  0.04 &  1.91 &  0.26 &  1.10  &  0.26 &  0.26 &  0.35 &  0.93 &  0.63 &  \textbf{0.93} &  \textbf{0.37} &  0.76 \\
\hline
\end{tabular}}
\end{table*}

\vspace{-0.3cm}
\subsection{Externally-Augmented Multi-Stage Fusion}

The \emph{externally-augmented} multi-stage fusion approach, shown in Fig. \ref{fig:fig9}(b), builds upon the concept of the self-augmented method but differs in how it incorporates the third score. While both the self-augmented three-score method and the externally-augmented version use scores from ECAPA-TDNN, AASIST, and RawGAT, their integration order varies. The self-augmented method fuses all three scores from the beginning, whereas the externally-augmented method begins with two initial scores (ASV from ECAPA-TDNN and CM from AASIST), and introduces the third score (CM from RawGAT) in a later stage, where it is combined with the classifier-derived SASV scores. The formulation for the externally-augmented multi-stage fusion is as follows:
% \vspace{-0.1cm}
\begin{equation}
\footnotesize
\begin{split}
    s = &\mathrm{C_{2}^{j}} \Bigl( \bigr[ \mathrm{C_{1}^{i}} \Bigl( \bigr[ s_{\texttt{ASV}}^\texttt{{ECAPA}},s_{\texttt{CM}}^\texttt{{AASIST}} \bigr] \Bigl), s_{\texttt{CM}}^\texttt{{RawGAT}} \bigr] \Bigl) \\
    &i, j \in \bigl\{ SVM, LR \bigl\}
\end{split}
\label{eq:mssf}
\end{equation}

Similar to the self-augmented multi-stage fusion, the externally-augmented multi-stage fusion again entails four variations of two-stage fusion according to classifiers chosed in the first and second stages. 

\section{Experimental Setup}

We evaluate our proposed multi-stage fusion systems on the logical access (LA) subset of the ASVspoof19 \cite{todisco2019asvspoof} database. The fusion models include an SVM classifier with a \emph{polynomial kernel} and an LR classifier with 10-fold cross-validation. Additionally, a Gaussian back-end classifier is employed for SASV. Three GMM models are trained for \emph{bonafide-target}, \emph{bonafide-nontarget}, and \emph{spoof-target} classes using development data. For each test utterance, three likelihood values are computed from the GMM models, and a single log-likelihood ratio score is derived using equation 1 from \cite{Todisco2018}.

The SASV2022 challenge baselines are based on the AASIST \cite{jung2022aasist} system, trained on the LA train partition of the ASVspoof2019 for CM, and the ECAPA-TDNN \cite{Desplanques_2020} model trained on VoxCeleb2 \cite{Chung_2018} dataset, for ASV. We use ASV and CM scores extracted from the ECAPA-TDNN and AASIST models, respectively.

We adopt two different kinds of evaluation metrics. Following the common strategy to compare SASV systems, the first one consists of three different \emph{equal error rates} (EERs) to measure discrimination of positive (target) class from different subsets of the negative class: the SV-EER, SPF-EER, and SASV-EER metrics use, respectively, non-targets, spoofing attacks, and a mix of nontargets and spoofing attacks as the negative class. Furthermore, we report recent \emph{architecture-agnostic detection cost function} (a-DCF) \cite{shim2024adcf}, which extends the classic DCF \cite{Doddington2000-NIST} used in ASV research to three-class evaluation. As opposed to reporting \emph{three} EERs, a-DCF gives \emph{one} number---the minimum normalized detection cost. We use the open-source implementation\footnote{\url{https://github.com/shimhz/a_dcf} (referred to on Mar 21, 2025)} with settings (target, nontarget, and spoof priors of 0.9, 0.05, and 0.05, respectively; and costs of target miss, nontarget false alarm, and spoof false alarm of 1, 10, and 20, respectively).

\vspace{-0.2cm}
\section{Results}
\vspace{-0.2cm}
\subsection{Overall Results}

Table \ref{table2} summarizes the experimental results of the proposed multi-stage score-level fusion frameworks for spoofing-aware speaker verification (SASV). The baseline system, using simple score summation, achieves a moderate SASV-EER of $1.71\%$ on the evaluation set. Single-stage fusion methods (a)-LR and (a)-SVM, which incorporate three scores, show improved performance with SASV-EERs of $1.55\%$ and $1.38\%$, and a-DCFs of $0.033$ and $0.029$, respectively.

The self-augmented multi-stage fusion methods demonstrate further improvements across different fusion paths. Notably, the (a)-(b)-(d)-(g) pathway, which uses three scores in a sequential setup, achieves the best performance with a SASV-EER of $1.30\%$ and a-DCF of $0.028$, underscoring the benefits of staged fusion. Fig. \ref{fig:fig10} visualizes score distributions for various features; while ECAPA-TDNN (Fig. \ref{fig:fig10}(a)) effectively separates target from non-target speakers, it overlaps with spoofed speech. In contrast, the best-performing system (Fig. \ref{fig:fig10}(c)) clearly distinguishes all three classes.

Externally-augmented multi-stage fusion variants, such as pathways (a)-(d)-(f)-(j) and (a)-(b)-(e)-(h), also outperform the 2-score fusion baseline, achieving SASV-EERs of $1.53\%$ with a-DCFs of $0.031$ and $0.035$. Overall, these findings highlight the effectiveness of multi-stage fusion in improving SASV system robustness.
\vspace{-0.25cm}

\subsection{Per-attack Results}
The individual results for each attack, presented in Table \ref{table3}, provide a thorough evaluation of system performance across different attack scenarios. The systems listed in this table are chosen based on their performance in Table \ref{table2}. The ECAPA-TDNN system exhibits high error rates across nearly all attack types. Conversely, the AASIST system performs exceptionally well for attacks A09, and it's the top performer for A10 and A16, but struggles with other attack types. Both the Baseline-B1 and (a)-SVM systems generally maintain low error rates, although there are slight increases observed for specific attacks. 

The multi-stage fusion methods, such as (a)-(c)-(e)-(i), (a)-(b)-(d)-(g), and (a)-(d)-(f)-(j), especially (a)-(b)-(d)-(g), demonstrate competitive performance, with the latter achieving notably low error rates. These findings emphasize the efficacy of multi-stage fusion techniques in bolstering system resilience across a variety of attack types.

\vspace{-0.2cm}
\section{Conclusion}
\vspace{-0.1cm}
In this study, we investigated a multi-stage score-level fusion to enhance the robustness of speaker verification systems against spoofing attacks. By integrating ASV and CM subsystems into an SASV framework, considerable improvements in detecting spoofed audio recordings were achieved. The multi-stage fusion methodology, which leverages SVM and LR classifiers, along with a new CM score from RawGAT, demonstrated robustness with an SASV-EER of 1.30\% and a-DCF of 0.028 for the SASV task.

\IEEEpeerreviewmaketitle

\IEEEpubidadjcol

% \begin{thebibliography}{1}

% \bibitem{paper1}
% M. Shand, S. Bryant, "IP Fast Reroute Framework", RFC 5714, 2010.

% \bibitem{paper2}
% A. Atlas, A. Zinin, "Basic Specification for IP Fast Reroute: Loop-Free Alternates", RFC 5286, 2008.

% \bibitem{paper3}
% S. Bryant, S. Previdi, M. Shand, "IP Fast Reroute Using Not-via Addresses", IETF Internet Draft, 2012.	

% \end{thebibliography}

% that's all folks

\begin{thebibliography}{99}

\bibitem{bai2021speaker}
Z. Bai and X.-L. Zhang, ``Speaker recognition based on deep learning: An overview,'' \textit{Neural Networks}, vol. 140, pp. 65–99, 2021.

\bibitem{Desplanques_2020}
B. Desplanques, J. Thienpondt, and K. Demuynck, ``ECAPA-TDNN: Emphasized channel attention, propagation and aggregation in tdnn based speaker verification,'' in \textit{Proc. Interspeech 2020}, ISCA, Oct. 2020. [Online]. Available: \url{http://dx.doi.org/10.21437/Interspeech.2020-2650}

\bibitem{8461375}
D. Snyder, D. Garcia-Romero, G. Sell, D. Povey, and S. Khudanpur, ``X-vectors: Robust dnn embeddings for speaker recognition,'' in \textit{Proc. 2018 IEEE International Conference on Acoustics, Speech and Signal Processing (ICASSP)}, 2018, pp. 5329–5333.

\bibitem{Shim2022}
H. Jin Shim, H. Tak, X. Liu, H.-S. Heo, J. Weon Jung, J. S. Chung, S.-W. Chung, H.-J. Yu, B.-J. Lee, M. Todisco, H. Delgado, K. A. Lee, M. Sahidullah, T. Kinnunen, and N. Evans, ``Baseline systems for the first spoofing-aware speaker verification challenge: Score and embedding fusion,'' in \textit{Proc. Odyssey 2022-The Speaker and Language Recognition Workshop}, 2022

\bibitem{Zhang2022}
Y. Zhang, G. Zhu, and Z. Duan, ``A probabilistic fusion framework for spoofing aware speaker verification,'' in \textit{Proc. The Speaker and
Language Recognition Workshop (Odyssey 2022)}, 2022, pp. 77–84.

\bibitem{kang22_interspeech}
W. Kang, M. J. Alam, and A. Fathan, ``End-to-end framework for spoof-aware speaker verification,'' in \textit{Proc. Interspeech 2022}, 2022, pp. 4362–4366.

\bibitem{Wu2022e}
H. Wu, L. Meng, J. Kang, J. Li, X. Li, X. Wu, H. Yi Lee, and H. Meng, ``Spoofing-aware speaker verification by multi-level fusion,'' in \textit{Proc. Interspeech 2022}, 4357-4361, doi: 10.21437/Interspeech.2022-920

\bibitem{Ge2022}
W. Ge, H. Tak, M. Todisco, and N. Evans, ``On the potential of jointly-optimised solutions to spoofing attack detection and automatic speaker verification,'' in \textit{Proc. IberSPEECH 2022}, 51-55, doi: 10.21437/IberSPEECH.2022-11.

\bibitem{Wang2022b}
X. Wang, X. Qin, Y. Wang, Y. Xu, and M. Li, ``The DKU-OPPO system for the 2022 spoofing-aware speaker verification challenge,'' in \textit{Proc. Interspeech 2022}, 4396-4400, doi: 10.21437/Interspeech.2022-11190

\bibitem{DBLP:journals/corr/abs-2106-15793}
X. Yao, S. Zhao, P. Xu, and J. Yang, ``Multi-source domain adaptation for object detection,'' in \textit{Proc. IEEE/CVF International Conference on Computer Vision}, 2021, pp. 3273–3282.

\bibitem{Atmaja_2020}
B. T. Atmaja and M. Akagi, ``Multitask learning and multistage fusion for dimensional audiovisual emotion recognition,'' in \textit{Proc. ICASSP 2020 - 2020 IEEE International Conference on Acoustics, Speech and Signal Processing (ICASSP)}, IEEE, May 2020. [Online]. Available: \url{http://dx.doi.org/10.1109/ICASSP40776.2020.9052916}

\bibitem{9522754}
S. Tulayakov, H. Sankaran, D. Mohan, S. Setlur, and V. Govindaraju, ``Multistage fusion of face matchers,'' in \textit{Proc. 2021 IEEE/CVF Conference on Computer Vision and Pattern Recognition Workshops (CVPRW)}, 2021, pp. 1444–1452.

\bibitem{todisco2019asvspoof}
M. Todisco, X. Wang, V. Vestman, M. Sahidullah, H. Delgado, A. Nautsch, J. Yamagishi, N. Evans, T. Kinnunen, and K. A. Lee, ``ASVspoof 2019: Future horizons in spoofed and fake audio detection,'' in \textit{Proc. Interspeech 2019}, pp. 1008–1012.

\bibitem{Todisco2018}
M. Todisco, H. Delgado, K. A. Lee, M. Sahidullah, N. Evans, T. Kinnunen, and J. Yamagishi, ``Integrated Presentation Attack Detection and Automatic Speaker Verification: Common Features and Gaussian Back-end Fusion,'' in \textit{Proc. Interspeech 2018}, 2018, pp. 77–81.

\bibitem{jung2022aasist}
J.-w. Jung, H.-S. Heo, H. Tak, H.-j. Shim, J. S. Chung, B.-J. Lee, H.-J. Yu, and N. Evans, ``AASIST: Audio anti-spoofing using integrated spectro-temporal graph attention networks,'' in \textit{Proc. ICASSP 2022 - 2022 IEEE International Conference on Acoustics, Speech and Signal Processing (ICASSP)}, IEEE, 2022, pp. 6367–6371.

\bibitem{Chung_2018}
J. S. Chung, A. Nagrani, and A. Zisserman, ``VoxCeleb2: Deep speaker recognition,'' in \textit{Proc. Interspeech 2018}, ISCA, Sep. 2018. [Online]. Available: \url{http://dx.doi.org/10.21437/Interspeech.2018-1299}

\bibitem{shim2024adcf}
H. Jin Shim, J. Weon Jung, T. Kinnunen, N. Evans, J.-F. Bonastre, and I. Lapidot, ``a-DCF: An architecture agnostic metric with application to spoofing-robust speaker verification,'' in \textit{Proc. The Speaker and Language Recognition Workshop (Odyssey 2024)}, pp. 158–164.

\bibitem{Doddington2000-NIST}
G. R. Doddington, M. A. Przybocki, A. F. Martin, and D. A. Reynolds, ``The NIST speaker recognition evaluation – overview, methodology, systems, results, perspective,'' \textit{Speech Communication}, vol. 31, no. 2, pp. 225–254, 2000.

\end{thebibliography}
\end{document}